\documentclass[runningheads]{llncs}
\usepackage{graphicx}
\usepackage[ruled, lines numbered, noend]{algorithm2e}
\SetKwProg{Task}{task}{}{end}
\SetKwInput{KwData}{Input}
\SetKwInput{KwEnsure}{Ensure}
\SetKwInput{KwResult}{Output}
\usepackage{amsmath,bm,amssymb}

\begin{document}

\title{Robust Task-Parallel Solution of the Triangular~Sylvester~Equation}
\author{Angelika Schwarz, Carl Christian Kjelgaard Mikkelsen}
\authorrunning{A. Schwarz, C. C. Kjelgaard Mikkelsen}
%
\institute{Department of Computing Science, Ume\r{a} University, Sweden\\
\email{\{angies,spock\}@cs.umu.se}}
\maketitle              
\begin{abstract}
The Bartels-Stewart algorithm is a standard approach to solving the dense Sylvester equation. It reduces the problem to the solution of the triangular Sylvester equation. The triangular Sylvester equation is solved with a variant of backward substitution. Backward substitution is prone to overflow. Overflow can be avoided by dynamic scaling of the solution matrix. An algorithm which prevents overflow is said to be robust. The standard library LAPACK contains the robust scalar sequential solver \texttt{dtrsyl}. This paper derives a robust, level-3 BLAS-based task-parallel solver. By adding overflow protection, our robust solver closes the gap between problems solvable by LAPACK and problems solvable by existing non-robust task-parallel solvers. We demonstrate that our robust solver achieves a similar performance as non-robust solvers.

\end{abstract}

\section{Introduction}\label{sec:introduction}

The Bartels-Stewart algorithm is a standard approach to solving the general Sylvester equation
\begin{equation}\label{eq:sylvester-dense}
\bm{A}\bm{X} + \bm{X}\bm{B} = \bm{C}
\end{equation}
where $\bm{A} \in \mathbb{R}^{m \times m}$, $\bm{B} \in \mathbb{R}^{n \times n}$ and $\bm{C} \in \mathbb{R}^{m \times n}$ are dense. It is well-known that (\ref{eq:sylvester-dense}) has a unique solution $\bm{X}\in \mathbb{R}^{m \times n}$ if and only if the eigenvalues $\lambda_i^{\bm{A}}$ of $\bm{A}$ and $\lambda_j^{\bm{B}}$ of $\bm{B}$ satisfy $\lambda_i^{\bm{A}} + \lambda_j ^{\bm{B}}\neq 0$ for all $i = 1, \hdots, m$ and $j = 1, \hdots, n$.

Sylvester equations occur in numerous applications including control and systems theory, signal processing and condition number estimation, see \cite{simoncini2016computational} for a summary.
The case of $\bm{B} = \bm{A}^T$ corresponds to the continuous-time Lyapunov matrix equation, which is central in the analysis of linear time-invariant dynamical systems.

The Bartels-Stewart algorithm solves (\ref{eq:sylvester-dense}) by reducing $\bm{A}$ and $\bm{B}$ to upper quasi-triangular form $\bm{\tilde{A}}$ and $\bm{\tilde{B}}$. This reduces the problem to the solution of the triangular Sylvester equation
\begin{equation}\label{eq:sylvester-triangular}
\bm{\tilde{A}}\bm{Y} + \bm{Y}\bm{\tilde{B}} = \bm{\tilde{C}}.
\end{equation}
During the solution of (\ref{eq:sylvester-triangular}) through a variant of backward substitution,
the entries of $\bm{Y}$ can exhibit growth, possibly exceeding the representable floating-point range. To avoid such an overflow, the LAPACK 3.7.0 routine \texttt{dtrsyl} uses a scaling factor $\alpha \in (0,1]$ to dynamically downscale the solution. We say that an algorithm is \emph{robust} if it cannot exceed the overflow threshold $\Omega$. With the scaling factor $\alpha$, the triangular Sylvester equation reads
$\bm{\tilde{A}}\bm{Y} + \bm{Y}\bm{\tilde{B}} = \alpha \bm{\tilde{C}}.$
This paper focuses on the robust solution of the triangular Sylvester equation and improves existing non-robust task-parallel implementations for~(\ref{eq:sylvester-triangular}) by adding protection against overflow. This closes the gap between the class of problems solvable by existing task-parallel solvers and the class of problems solvable by LAPACK. Consequently, more problems can be solved efficiently in parallel through the Bartels-Stewart method.

\begin{algorithm}[t]
\caption{Triangular Sylvester Equation Solver}\label{algo:NonRobustSyl}
\KwData{$\bm{\tilde{A}}$, $\bm{\tilde{B}}$ as in (\ref{eq:a-shape}), conformally partitioned $\bm{\tilde{C}}$ as in (\ref{eq:c-shape}).}
\KwResult{$\bm{Y} \in \mathbb{R}^{m \times n}$ such that $\bm{\tilde{A}} \bm{Y} + \bm{Y}\bm{\tilde{B}} = \bm{\tilde{C}}$.}
\SetKwProg{Fn}{}{}{}
  $\bm{Y} \gets \bm{\tilde{C}}$\;
  \For{$\ell \gets 1, 2, \hdots{}, q$}{
    \For{$k \gets p, p - 1, \hdots, 1$}{
      Solve $\bm{\tilde{A}}_{kk} \bm{Z} + \bm{Z} \bm{\tilde{B}}_{\ell \ell} = \bm{Y}_{k \ell}$ for $\bm{Z}$\;
      $\bm{Y}_{k \ell} \gets \bm{Z}$\;
      $\bm{Y}_{1:k-1,\ell} \gets \bm{Y}_{1:k-1,\ell} - \bm{\tilde{A}}_{1:k-1,k}\bm{Y}_{k \ell}$\;
      $\bm{Y}_{k,\ell+1:q} \gets \bm{Y}_{k,\ell+1:q} - \bm{Y}_{k \ell} \bm{\tilde{B}}_{\ell, \ell+1:q}$\;
    }
  }
  \Return $\bm{Y}$\;
\end{algorithm}

We now describe the Bartels-Stewart algorithm for solving~(\ref{eq:sylvester-dense}), see \cite{bartels1972,golub1996matrix,simoncini2016computational}. The algorithm computes the real Schur decompositions of $\bm{A}$ and $\bm{B}$
\begin{equation}\label{eq:schur}
\bm{A} = \bm{U} \bm{\tilde{A}} \bm{U}^T, \quad \bm{B}  = \bm{V}  \bm{\tilde{B}}\bm{V}^T
\end{equation}
using orthogonal transformations $\bm{U}$ and $\bm{V}$. Using (\ref{eq:schur}), the general Sylvester equation~(\ref{eq:sylvester-dense}) is transformed into the triangular Sylvester equation
\begin{equation*}
\bm{\tilde{A}} \bm{Y} + \bm{Y} \bm{\tilde{B}} = \bm{\tilde{C}}, \quad \bm{\tilde{C}} = \bm{U}^T \bm{C} \bm{V}, \quad  \bm{Y} = \bm{U}^T \bm{X} \bm{V}.
\end{equation*}
The solution of the original system (\ref{eq:sylvester-dense}) is given by $\bm{X} = \bm{U} \bm{Y} \bm{V}^T$.
The real Schur forms $\bm{\tilde{A}}$ and $\bm{\tilde{B}}$ attain the shapes
\begin{equation}\label{eq:a-shape}
\bm{\tilde{A}} = 
\left[\begin{array}{cccc}
\bm{\tilde{A}}_{11} & \bm{\tilde{A}}_{12} & \hdots  & \bm{\tilde{A}}_{1p}\\
                    & \bm{\tilde{A}}_{22} &         & \bm{\tilde{A}}_{2p}\\
                    &                     & \ddots  & \vdots\\
                    &                     &         & \bm{\tilde{A}}_{pp}
\end{array}\right] \in \mathbb{R}^{m \times m}, \quad
\bm{\tilde{B}} = 
\left[\begin{array}{cccc}
\bm{\tilde{B}}_{11}      & \bm{\tilde{B}}_{12} & \hdots & \bm{\tilde{B}}_{1q}\\
                 & \bm{\tilde{B}}_{22} & \hdots & \bm{\tilde{B}}_{2q}\\
                 &             & \ddots & \vdots \\
                 &             &        & \bm{\tilde{B}}_{qq}
\end{array}\right] \in \mathbb{R}^{n \times n},
\end{equation}
where the diagonal blocks $\bm{\tilde{A}}_{kk}$ and $\bm{\tilde{B}}_{\ell \ell}$ are either 1-by-1 or 2-by-2. The right-hand side $\bm{\tilde{C}} = \bm{U}^T \bm{C} \bm{V}$ is partitioned conformally
\begin{equation}\label{eq:c-shape}
\bm{\tilde{C}} = 
\bm{U}^T \bm{C} \bm{V} =
\left[\begin{array}{ccc}
\bm{\tilde{C}}_{11}  & \hdots  & \bm{\tilde{C}}_{1q}\\
\vdots               &  \ddots & \vdots \\
\bm{\tilde{C}}_{p1}  &  \hdots & \bm{\tilde{C}}_{pq}
\end{array}\right] \in \mathbb{R}^{m \times n}.
\end{equation}
Adopting a block perspective, (\ref{eq:sylvester-triangular}) reads
\begin{equation}\label{eq:block-wise-sylvester}
\bm{\tilde{A}}_{kk} \bm{Y}_{k \ell} + \bm{Y}_{k \ell} \bm{\tilde{B}}_{\ell \ell} = \bm{\tilde{C}}_{k \ell} - \left(\sum \limits_{i = 1}^{k-1} \bm{\tilde{A}}_{ki} \bm{Y}_{i \ell} + \sum \limits_{j = \ell+1}^{q} \bm{Y}_{ik} \bm{\tilde{B}}_{kj}\right)
\end{equation}
for all blocks $k = 1, \hdots, p$ and $\ell = 1, \hdots q$. A straight-forward implementation of (\ref{eq:block-wise-sylvester}) is Algorithm~\ref{algo:NonRobustSyl}. The algorithm starts at the bottom left corner ($k=p,\ell=1$) and processes the block columns bottom-up from left to right. The flop count of the algorithm approximately corresponds to two backward substitutions and amounts to $\mathcal{O}(m^2n+mn^2)$ flops.

The stability of Algorithm~\ref{algo:NonRobustSyl} has been summarized in Higham~\cite{higham2002accuracy}. In essence, the algorithm inherits the stability from backward substitution provided that the small Sylvester equations (line 4) can be solved stably.

The rest of this paper is structured as follows. In Section~\ref{sec:robust} we formalize the definition of a robust algorithm and derive a new robust algorithm for solving the triangular Sylvester equation. Section~\ref{sec:test-setup} describes the execution environment used in the numerical experiments in Section~\ref{sec:performance-results}. Section~\ref{sec:conclusion} summarizes the results.

\section{Robust Algorithms for Triangular Sylvester Equations}\label{sec:robust}

In this section, we address the robust solution of the triangular Sylvester equation. 
The goal is to dynamically compute a scaling factor $\alpha \in (0,1]$ such that the solution $\bm{Y}$ of the scaled triangular Sylvester equation
\begin{align}\label{eq:sylvester-triangular-robust-repeated}
\bm{\tilde{A}}\bm{Y} + \bm{Y}\bm{\tilde{B}} = \alpha \bm{\tilde{C}}
\end{align}
can be obtained without ever exceeding the overflow threshold $\Omega > 0$. We derive two robust algorithms for solving (\ref{eq:sylvester-triangular-robust-repeated}). The first scalar algorithm can be viewed as an enhancement of LAPACK's \texttt{dtrsyl} by adding overflow protection to the linear updates. The second tiled algorithm redesigns the first algorithm such that most of the computation is executed as matrix-matrix multiplications.

\subsection{Scalar Robust Algorithm}
The central building block for adding robustness is \textsc{ProtectUpdate}, introduced by Kjelgaard Mikkelsen and Karlsson in~\cite{mikkelsen2017nlafet}. \textsc{ProtectUpdate} computes a scaling factor $\zeta\in (0,1]$ such that the matrix update $\zeta \bm{C} - \bm{A}(\zeta \bm{Y})$ cannot overflow. \textsc{ProtectUpdate} uses the upper bounds $||\bm{C}||_\infty$, $||\bm{A}||_\infty$ and $||\bm{Y}||_\infty$ to evaluate the maximum growth possible in the update. Provided that $||\bm{C}||_\infty \leq \Omega$, $||\bm{A}||_\infty \leq \Omega$ and $||\bm{Y}||_\infty \leq \Omega$, \textsc{ProtectUpdate} computes a scaling factor $\zeta$ such that $\zeta (||\bm{C}||_\infty + ||\bm{A}||_\infty ||\bm{Y}||_\infty) \leq \Omega$. 

We use \textsc{ProtectUpdate} to protect the left and the right updates in the triangular Sylvester equation. We protect right updates by applying the scaling factor as $\zeta \bm{C} - \bm{\tilde{A}}(\zeta\bm{Y})$. We protect left updates by appling the scaling factor as $\zeta \bm{C} - (\zeta \bm{Y}) \bm{\tilde{B}}$.

A solver for the triangular Sylvester equation requires the solution of small Sylvester equations $\bm{\tilde{A}}_{kk} \bm{Z} + \bm{Z} \bm{\tilde{B}}_{\ell \ell} = \beta \bm{Y}_{k \ell}$, $\beta \in (0,1]$. Since $\bm{\tilde{A}}_{kk}$ and $\bm{\tilde{B}}_{\ell \ell}$ are at most 2-by-2, these small Sylvester equations can be converted into linear systems of size at most 4-by-4, see~\cite{bartels1972}. We solve these linear systems robustly through Gaussian elimination with complete pivoting. This process requires linear updates and divisions to be executed robustly. We use \textsc{ProtectUpdate} for the small linear updates and guard divisions with \textsc{ProtectDivision} from~\cite{mikkelsen2017nlafet}.

\begin{algorithm}[th]
\caption{Robust Triangular Sylvester Equation Solver}\label{algo:RobustSyl}
\KwData{$\bm{\tilde{A}}$, $\bm{\tilde{B}}$ as in (\ref{eq:a-shape}) with $||\bm{\tilde{A}}_{ij}||_\infty \leq \Omega$, $||\bm{\tilde{B}}_{k\ell}||_\infty \leq \Omega$, conformally partitioned $\bm{\tilde{C}}$ as in (\ref{eq:c-shape}) with $||\bm{\tilde{C}}_{i\ell}||_\infty \leq \Omega$.}
\KwResult{$\alpha \in (0,1]$, $\bm{Y} \in \mathbb{R}^{m \times n}$ such that $\bm{\tilde{A}} \bm{Y} + \bm{Y}\bm{\tilde{B}} = \alpha \bm{\tilde{C}}$.}
\KwEnsure{ $||\bm{Y}_{k\ell}||_\infty \leq \Omega$ for $\bm{Y}$ partitioned analogously to $\bm{\tilde{C}}$ as in (\ref{eq:c-shape}).}
\SetKwProg{Fn}{}{}{}
\Fn{\textsc{RobustSyl}($\bm{\tilde{A}}$, $\bm{\tilde{B}}$, $\bm{\tilde{C}}$)}{
  $\bm{Y} \gets \bm{\tilde{C}}$; $\alpha \gets 1$\;
  \For{$\ell \gets 1, 2, \hdots{}, q$}{
    \For{$k \gets p, p - 1, \hdots, 1$}{
      Solve robustly $\bm{\tilde{A}}_{kk} \bm{Z} + \bm{Z} \bm{\tilde{B}}_{\ell \ell} = \beta \bm{Y}_{k \ell}$ for $\beta$, $\bm{Z}$\;
      $\bm{Y} \gets \beta \bm{Y}$\;
      $\bm{Y}_{k \ell} \gets \bm{Z}$\;
      $\gamma_1 \gets \textsc{ProtectUpdate}(||\bm{Y}_{1:k-1,\ell}||_\infty, ||\bm{\tilde{A}}_{1:k-1,k}||_\infty, ||\bm{Y}_{k \ell}||_\infty)$\;
      $\bm{Y} \gets \gamma_1 \bm{Y}$\;
      $\bm{Y}_{1:k-1,\ell} \gets \bm{Y}_{1:k-1,\ell} - \bm{\tilde{A}}_{1:k-1,k}\bm{Y}_{k \ell}$\;
      $\gamma_2 \gets \textsc{ProtectUpdate}(\newline \hspace*{1.2cm} \max \limits_{\ell +1\leq j \leq q} \{||\bm{Y}_{k,j}||_\infty\}, ||\bm{Y}_{k \ell}||_\infty, \max \limits_{\ell+1 \leq j \leq q\}}\{||\bm{\tilde{B}}_{\ell, j}||_\infty\})$\;
      $\bm{Y} \gets \gamma_2 \bm{Y}$\;
      $\bm{Y}_{k,\ell+1:q} \gets \bm{Y}_{k,\ell+1:q} - \bm{Y}_{k \ell} \bm{\tilde{B}}_{\ell, j+1:q}$\;
      $\alpha \gets \alpha \beta \gamma_1 \gamma_2$\;
    }
  }
  \Return $\alpha$, $\bm{Y}$\;
}
\end{algorithm}

Algorithm~\ref{algo:RobustSyl} \textsc{RobustSyl} adds overflow protection to Algorithm~\ref{algo:NonRobustSyl}. \textsc{RobustSyl} is dominated by level-2 BLAS-like thin linear updates. The next section develops a solver, which relies on efficient level-3 BLAS operations and uses \textsc{RobustSyl} as a basic building block.

\subsection{Tiled Robust Algorithm}
Solvers for the triangular Sylvester equation can be expressed as tiled algorithms~\cite{quintana2003formal} such that most of the computation corresponds to matrix-matrix multiplications. For this purpose, the matrices are partitioned into conforming, contiguous submatrices, so-called \emph{tiles}. In order to decouple the tiles, the global scaling factor $\alpha$ is replaced with local scaling factors, one per tile of $\bm{Y}$. The association of a tile with a scaling factor leads to \emph{augmented tiles}.

\begin{definition}
An augmented tile $\langle \alpha, \bm{X}\rangle$ consists of a scalar $\alpha \in (0,1]$ and a matrix $\bm{X} \in \mathbb{R}^{m \times n}$ and represents the scaled matrix $\bm{Y} = \alpha^{-1} \bm{X}$. We say that two augmented tiles $\langle \alpha, \bm{X}\rangle$ and $\langle \beta, \bm{Y}\rangle$ are equivalent and we write $\langle \alpha, \bm{X}\rangle = \langle \beta, \bm{Y}\rangle$ if and only if $ \alpha^{-1} \bm{X} = \beta^{-1} \bm{Y}$.
\end{definition}

\begin{definition}
We say that two augmented tiles  $\langle \alpha, \bm{X}\rangle$ and $\langle \beta, \bm{Y}\rangle$ are consistently scaled if $\alpha = \beta$.
\end{definition}

The idea of associating a scaling factor with a vector was introduced by Kjelgaard Mikkelsen and Karlsson~\cite{mikkelsen2017nlafet,mikkelsen2017blocked} who use augmented vectors to represent scaled vectors. We generalize their definition and associate a scaling factor with a tile. Their definition of consistently scaled vectors generalizes likewise to consistently scaled tiles.

Let $\bm{\tilde{A}}$ and $\bm{\tilde{B}}$ be as in (\ref{eq:a-shape}). A partitioning of $\bm{\tilde{A}}$ into $\rm{M}$-by-$\rm{M}$ tiles and $\bm{\tilde{B}}$ into $\rm{N}$-by-$\rm{N}$ tiles such that 2-by-2 blocks on the diagonals are not split and the diagonal tiles are square induces a partitioning of $\bm{Y}$ and $\bm{\tilde{C}}$. We then solve the tiled equations
\begin{align}\label{eq:tiled}
\begin{split}
\bm{\tilde{A}}_{kk} & (\alpha_{k \ell}^{-1}\bm{Y}_{k \ell}) + (\alpha_{k \ell}^{-1}\bm{Y}_{k \ell}) \bm{\tilde{B}}_{\ell \ell} = \\ &\bm{\tilde{C}}_{k \ell} - \left(\sum \limits_{i = 1}^{k-1} \bm{\tilde{A}}_{ki} (\alpha_{i \ell}^{-1}\bm{Y}_{i \ell}) + \sum \limits_{j = \ell+1}^{q} (\alpha_{ik}^{-1}\bm{Y}_{ik}) \bm{\tilde{B}}_{kj}\right), 
\end{split}
\end{align}
\begin{algorithm}[t]
\caption{Robust Linear Tile Update}\label{algo:TileUpdate}
\KwData{$\bm{A} \in \mathbb{R}^{m \times k}$ with $||\bm{A}||_\infty \leq \Omega$ , $\bm{B} \in \mathbb{R}^{k \times n}$ with $||\bm{B}||_\infty \leq \Omega$ , $\bm{C} \in \mathbb{R}^{m \times n}$ with $||\bm{C}||_\infty \leq \Omega$ and scalars $\alpha$, $\beta$, $\gamma \in (0,1]$.}
\KwResult{$\bm{D} \in \mathbb{R}^{m \times n}$ and $\delta \in (0,1]$ such that $(\delta^{-1}\bm{D}) \gets (\gamma^{-1} \bm{C}) - (\alpha^{-1} \bm{A}) (\beta^{-1} \bm{B}) $.}
\SetKwProg{Fn}{}{}{}
\Fn{\textsc{RobustUpdate}$(\langle \gamma, \bm{C} \rangle$, $\langle \alpha, \bm{A}\rangle$, $\langle \beta, \bm{B} \rangle )$}{
  $\eta \gets \min\{\gamma, \alpha, \beta\}$\;
  $\zeta \gets \textsc{ProtectUpdate}((\eta/\gamma)||\bm{C}||_{\infty}), (\eta/\alpha)||\bm{A}||_{\infty}, (\eta/\beta)||\bm{B}||_{\infty}) $\;
  $\delta \gets \eta \zeta$\;
  $\bm{D} \gets (\delta/\gamma) \bm{C} - [(\delta/\alpha)\bm{A}] \, [(\delta/\beta) \bm{B}]$\;
  \Return $\langle \delta, \bm{D} \rangle$\;
}
\end{algorithm}
$k = 1, \hdots, \rm{M},  \ell = 1, \hdots, \rm{N}$ without explicitly forming any of the products $\alpha_{kl}^{-1}\bm{Y}_{kl}$, $\alpha_{i\ell}^{-1}\bm{Y}_{i\ell}$ and $\alpha_{ik}^{-1}\bm{Y}_{ik}$. Note that (\ref{eq:tiled}) is structurally identical to (\ref{eq:block-wise-sylvester}). The solution of (\ref{eq:tiled}) requires augmented tiles to be updated. Algorithm~\ref{algo:TileUpdate} \textsc{RobustUpdate} executes such an update robustly. Combining \textsc{RobustUpdate} and \textsc{RobustSyl} leads to Algorithm~\ref{algo:tiled}, which solves (\ref{eq:sylvester-triangular-robust-repeated}) in a tiled fashion. The global scaling factor $\alpha$ corresponds to the smallest of the local scaling factors $\alpha_{k\ell}$. The solution $\bm{Y}$ is obtained by scaling the tiles consistently.

Algorithm~\ref{algo:tiled} can be parallelized with tasks. Each function call corresponds to a task. \textsc{RobustSyl} on $\bm{Y}_{k\ell}$ has outgoing dependences to \textsc{RobustUpdate} modifying $\bm{Y}_{il}$, $i = 1, \hdots, k - 1$ and $\bm{Y}_{kj}$, $j = \ell + 1, \hdots \rm{N}$.
The incoming dependences of a \textsc{RobustSyl} task on $\bm{Y}_{k\ell}$ are satisfied when all updates modifying $\bm{Y}_{k\ell}$ have been completed. The updates to $\bm{Y}_{k\ell}$ require exclusive write access, which we achieve with a lock.

\textsc{RobustUpdate} tasks rely on upper bounds $||\bm{\tilde{A}}_{ik}||_\infty$ and $||\bm{\tilde{B}}_{\ell j}||_\infty$. Since the computation of norms is expensive, the matrix norms are precomputed and recorded. This is realized through perfectly parallel \textsc{Bound} tasks. To limit the amount of dependences to be handled by the runtime system, a synchronization point separates this preprocessing step and \textsc{RobustSyl}/\textsc{RobustUpdate} tasks.

Another synchronization point precedes the consistency scaling. The local scaling factors are reduced sequentially to the global scaling factor $\alpha$. The consistency scaling of the tiles is executed with independent \textsc{Scale} tasks.

\begin{algorithm}[t]
\caption{Tiled Robust Triangular Sylvester Equation Solver}\label{algo:tiled}
\KwData{$\tilde{\bm{A}}$, $\tilde{\bm{B}}$, $\tilde{\bm{C}}$ as in (\ref{eq:tiled}) where $||\bm{\tilde{A}}_{ij}||_\infty \leq \Omega$, $||\bm{\tilde{B}}_{k \ell}||_\infty \leq \Omega$, $||\bm{\tilde{C}}_{i\ell}||_\infty \leq \Omega$.}
\KwResult{$\alpha \in (0,1]$, $\bm{Y} \in \mathbb{R}^{m \times n}$ such that $\bm{\tilde{A}} \bm{Y} + \bm{Y}\bm{\tilde{B}} = \alpha \bm{\tilde{C}}$.}
\KwEnsure{For $\bm{Y}$ as in (\ref{eq:tiled}) where $||\bm{Y}_{k\ell}||_\infty \leq \Omega$.}
\SetKwProg{Fn}{}{}{}
\Fn{\textsc{drsylv}$(\bm{\tilde{A}}, \bm{\tilde{B}}, \bm{\tilde{C}})$}{
\For{$k \gets \rm{M}, \rm{M} - 1, \hdots, 1$}{
  \For{$\ell \gets 1, \hdots{}, \rm{N}$}{
$\langle \alpha_{k\ell}, \bm{Y}_{k\ell} \rangle \gets \langle 1,  \bm{\tilde{C}}_{k\ell}\rangle$\;
}}

\For{$k \gets \rm{M}, \rm{M} - 1, \hdots, 1$}{
  \For{$\ell \gets 1, \hdots{}, \rm{N}$}{   
      $\alpha_{kl}, \bm{Y}_{k\ell} \gets \textsc{RobustSyl}(\bm{\tilde{A}}_{kk}, \bm{\tilde{B}}_{\ell\ell}, \bm{Y}_{k\ell})$\;      
    \For{$i \gets k - 1, k - 2, \hdots, 1$} {
        $\langle \alpha_{i\ell}, \bm{Y}_{i\ell}\rangle \gets \textsc{RobustUpdate}(\langle \alpha_{i\ell}, \bm{Y}_{i\ell}\rangle, \langle 1, \bm{\tilde{A}}_{ik}\rangle, \langle \alpha_{k\ell}, \bm{Y}_{k\ell} \rangle)$\;
    }
    \For{$j \gets \ell + 1, \ell + 2, \hdots, \rm{N}$}{
        \mbox{$\langle \alpha_{kj}, \bm{Y}_{kj} \rangle \gets \textsc{RobustUpdate}(\langle \alpha_{kj}, \bm{Y}_{kj} \rangle, \langle \alpha_{k \ell}, \bm{Y}_{k \ell} \rangle, \langle 1, \bm{\tilde{B}}_{\ell j}\rangle)$}\;
    }
  }
}
$\alpha \gets \min \limits_{1 \leq k \leq \rm{M}, 1 \leq \ell \leq \rm{N}} \{\alpha_{k \ell}\}$ \tcp*{Compute global scaling factor}
\For{$k \gets \rm{M}, \rm{M} - 1, \hdots, 1$}{
  \For{$\ell \gets 1, \hdots{}, \rm{N}$}{
    $\bm{Y}_{k \ell} \gets \left(\alpha / \alpha_{k \ell}\right) \bm{Y}_{k \ell}$ \tcp*{Consistency scaling}
  }
}
\Return $\alpha, \bm{Y}$\;
}
\end{algorithm}

\section{Experiment Setup}\label{sec:test-setup}
This section describes the setup of the numerical experiments. We specify the hardware, the solvers and their configuration, and the matrices of the triangular Sylvester equation.

\paragraph{Execution Environment}
The experiments were executed on an Intel Xeon E5-2690v4 (``Broadwell'') node with 28 cores arranged in two NUMA islands with 14 cores each. The theoretical peak performance in double-precision arithmetic is 41.6 GFLOPS/s for one core and 1164.8 GFLOPS/s for a full node. In the STREAM triad benchmark the single core memory bandwidth was measured at 19 MB/s; the full node reached 123 MB/s.

We use the GNU compiler 6.4.0 and link against single-threaded OpenBLAS 0.2.20 and LAPACK 3.7.0. The compiler optimization flags are \texttt{-O2 -xHost}. We forbid migration of threads and fill one NUMA island with threads before assigning threads to the second NUMA island by setting \texttt{OMP\_PROC\_BIND} to \texttt{close}.

\paragraph{Software}

This section describes the routines used in the numerical experiments. The first two routines are non-robust, i.e., the routines solve $\bm{\tilde{A}}\bm{Y} + \bm{Y}\bm{\tilde{B}} = \bm{\tilde{C}}$.
\begin{itemize}
\item \texttt{FLA\_Sylv}. The \texttt{libflame} version~5.1.0-58~\cite{libflame,quintana2003formal} solver partitions the problem into tiles and executes the linear updates as matrix-matrix multiplications.
\item \texttt{FLASH\_Sylv}. This \texttt{libflame} routine is the supermatrix version of \texttt{FLA\_Sylv} and introduces task parallelism.
\end{itemize}
The following three routines are robust and solve $\bm{\tilde{A}}\bm{Y} + \bm{Y}\bm{\tilde{B}} = \alpha \bm{\tilde{C}}$.
\begin{itemize}
\item \texttt{dtrsyl}. The LAPACK 3.7.0 routine realizes overflow protection with a global scaling factor. Any scaling events triggers scaling of the \emph{entire} matrix $\bm{Y}$.
\item \texttt{recsy}. The Fortran library~release 2009-12-28~\cite{jonsson2002part1} offers recursive blocked solvers for a variety of Sylvester and Lyapunov equations. If overflow protection is triggered at some recursion level, the required scaling is propagated. 
\item \texttt{drsylv}. Our solver can be viewed as a robust version of \texttt{FLASH\_Sylv}. Due to the usage of local scaling factors, our solver exhibits the same degree of parallelism as \texttt{FLASH\_Sylv}.
\end{itemize}

\paragraph{Test Matrices}
We design the system matrices such that the growth during the solve is controlled. This allows us to (a) design systems that the non-robust solvers must be able to solve and (b) examine the cost of robustness by increasing the growth and, in turn, the amount of scaling necessary.

The matrix $\bm{\tilde{C}}$ and the upper triangular part of $\bm{\tilde{A}}$ and $\bm{\tilde{B}}$ are filled with ones. The diagonal blocks are set to $\bm{\tilde{A}}_{ii} = \mu \bm{T}_{ii}$ and $\bm{\tilde{B}}_{ii} = \nu \bm{T}_{ii}$, where $\bm{T}_{ii}$ is either the 1-by-1 block given by $t_{ii} = 1$ or the 2-by-2 block given by
\begin{align*}\bm{T}_{ii} = \left[\begin{array}{cc}
1 & 1\\
-1 & 1
\end{array}\right].
\end{align*}
The magnitude of the diagonal entries of $\bm{\tilde{A}}$ and $\bm{\tilde{B}}$ is controlled by $\mu$ and $\nu$. This also holds for 2-by-2 blocks, which encode a complex conjugate pair of eigenvalues. The 2-by-2 blocks cannot be reduced to triangular form using a real similarity transformation. A unitary transformation, however, can transform the matrix into triangular shape. As an example, consider the transformation $\bm{Q}^H \bm{\tilde{A}} \bm{Q}$ into triangular shape for the 5-by-5 matrix
\begin{align*}
\bm{\tilde{A}} = 
\left[
\begin{array}{ccccc}
\mu & 1   & 1  & 1   & 1\\
    & \mu & 1  & 1   & 1 \\
    &     &\mu & \mu & 1 \\
    &     &-\mu& \mu & 1 \\
    &     &    &     & \mu
\end{array}
\right], \quad
\bm{Q}^H \bm{\tilde{A}} \bm{Q} = 
\left[
\begin{array}{ccccc}
\mu    & 1            & \frac{-1+i}{\sqrt{2}} & \frac{1-i}{\sqrt{2}} & 1\\
       & \mu          & \frac{-1+i}{\sqrt{2}} & \frac{1-i}{\sqrt{2}} & 1\\
       &              & \mu (1+i)             & 0                    &\frac{-1-i}{ \sqrt{2}}\\
       &              &                       & \mu(1-i)             & \frac{1+i}{\sqrt{2}}\\
       &              &                       &                      & \mu (1+i)
\end{array}
\right].
\end{align*}
It is clear that a small value of $\mu$ introduces growth during the backward substitution. Hence, the choice of $\mu$ and $\nu$ control the growth during the solve.

\paragraph{Tuning}
The tile sizes for \texttt{FLASH\_Sylv} and \texttt{drsylv} were tuned using $\mu =m$ and $\nu = n$. For each core count, a sweep over [100, 612] with a step size of 32 was evaluated three times and the tile size with the best median runtime was chosen.

\paragraph{Reliability}
We extend the relative residual defined by Higham~\cite{higham2002accuracy} with the scaling factor $\alpha$ and evaluate
\begin{equation}\label{eq:relative-residual}
\frac{||\bm{R}||_F}{(||\bm{\tilde{A}}||_F + ||\bm{\tilde{B}}||_F)||\bm{Y}||_F + ||\alpha \bm{\tilde{C}}||_F},
\end{equation}
where $\bm{R} \gets \alpha \bm{\tilde{C}} - (\bm{\tilde{A}}\bm{Y} + \bm{Y}\bm{\tilde{B}})$. We report the median runtime of three runs.

\section{Performance Results}\label{sec:performance-results}

This section presents three sets of performance results. First, the five solvers are executed in sequential mode. Second, the parallel scalability is analyzed. Third, the cost of robustness is investigated.

\paragraph{Sequential Comparison without Numerical Scaling}

\begin{figure}[t]
  \centering
  \begin{minipage}[t]{.34\textwidth}
    \centering
    \includegraphics[width=\linewidth]{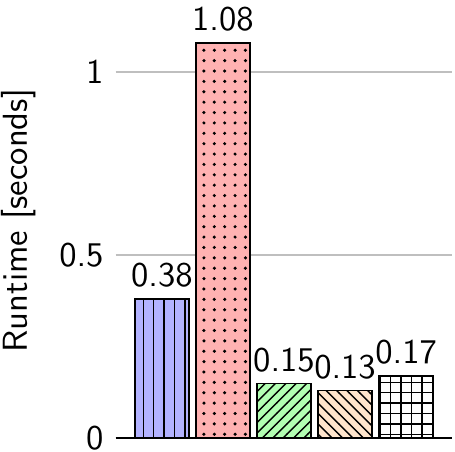}
    $m = n = 1000$
  \end{minipage}
  \begin{minipage}[t]{.34\textwidth}
    \centering
    \includegraphics[width=\linewidth]{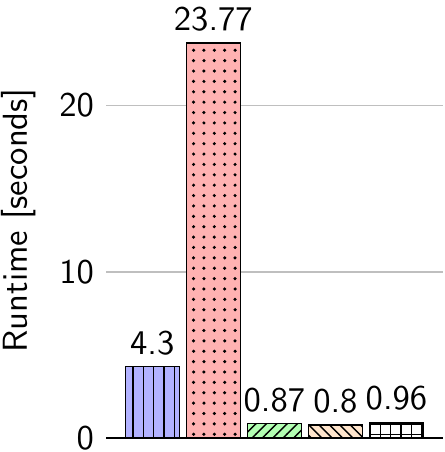}
    $m = n = 2000$
  \end{minipage}
  \begin{minipage}[t]{.28\textwidth}
    \includegraphics[width=.8\linewidth]{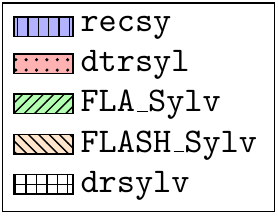}
    \vfill
  \end{minipage}
  \caption{Sequential runtime comparison on systems that do not require scaling.}\label{fig:sequential-comparison}
\end{figure}

Figure~\ref{fig:sequential-comparison} compares the solvers using $\mu = m$ and $\nu = n$. With this choice, dynamic downscaling of $\bm{Y}$ is not necessary. Our solver \texttt{drsylv} is slightly slower than \texttt{FLA\_Sylv} and \texttt{FLASH\_Sylv}, possibly because the overhead from robustness cannot be amortized. Since the gap between \texttt{recsy} and \texttt{dtrsyl} on the one hand and \texttt{FLA\_Sylv}, \texttt{FLASH\_Sylv} and \texttt{drsylv} on the other hand grows with increasing matrix sizes, we restrict the parallel experiments to a comparison between \texttt{drsylv} and \texttt{FLASH\_Sylv}.

\begin{figure}[t]
  \begin{minipage}{.5\textwidth}
    \centering
    \includegraphics[width=\linewidth]{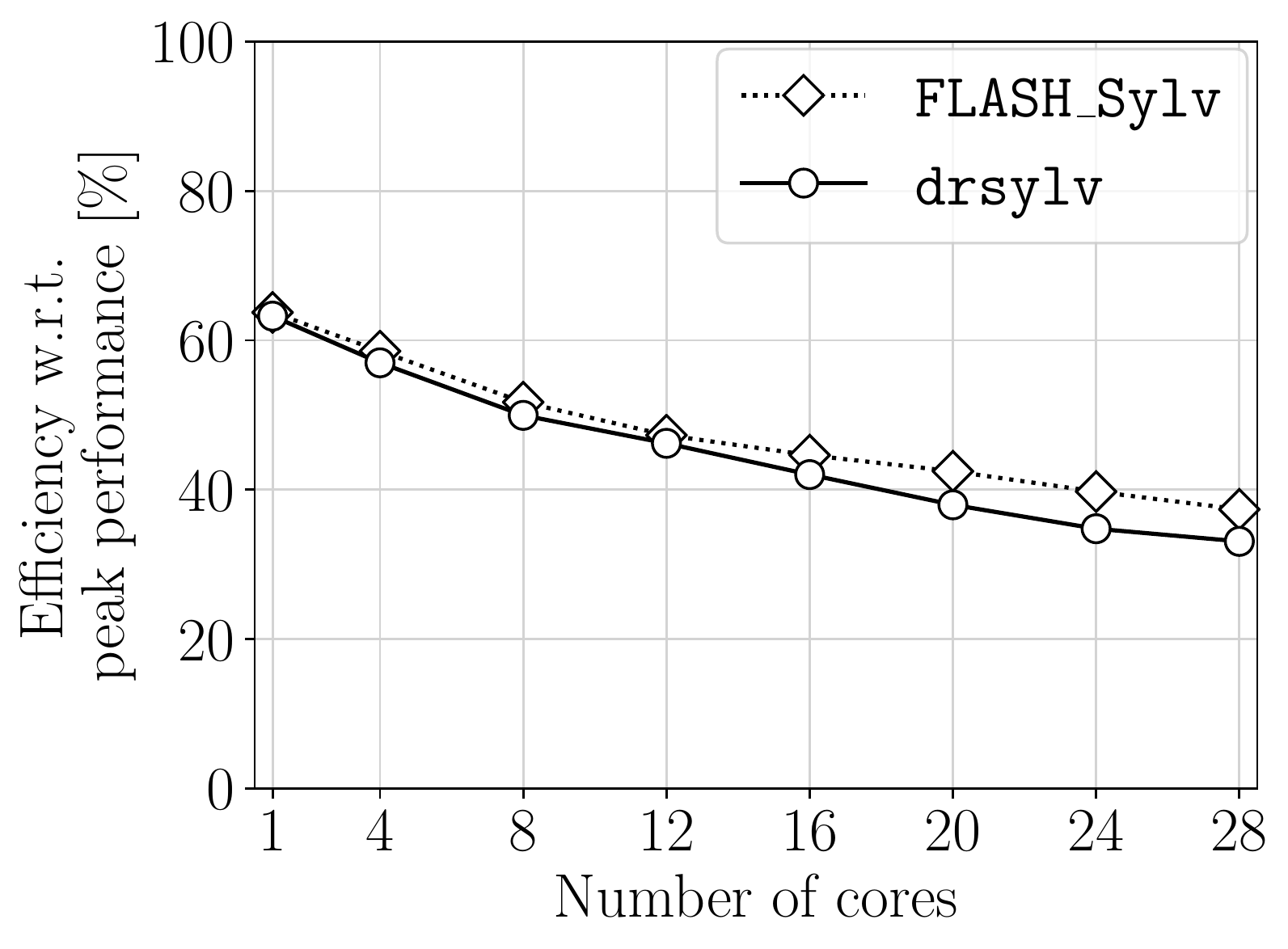}
    $m = n = 10000$
  \end{minipage}
  \begin{minipage}{.5\textwidth}
    \centering
    \includegraphics[width=\linewidth]{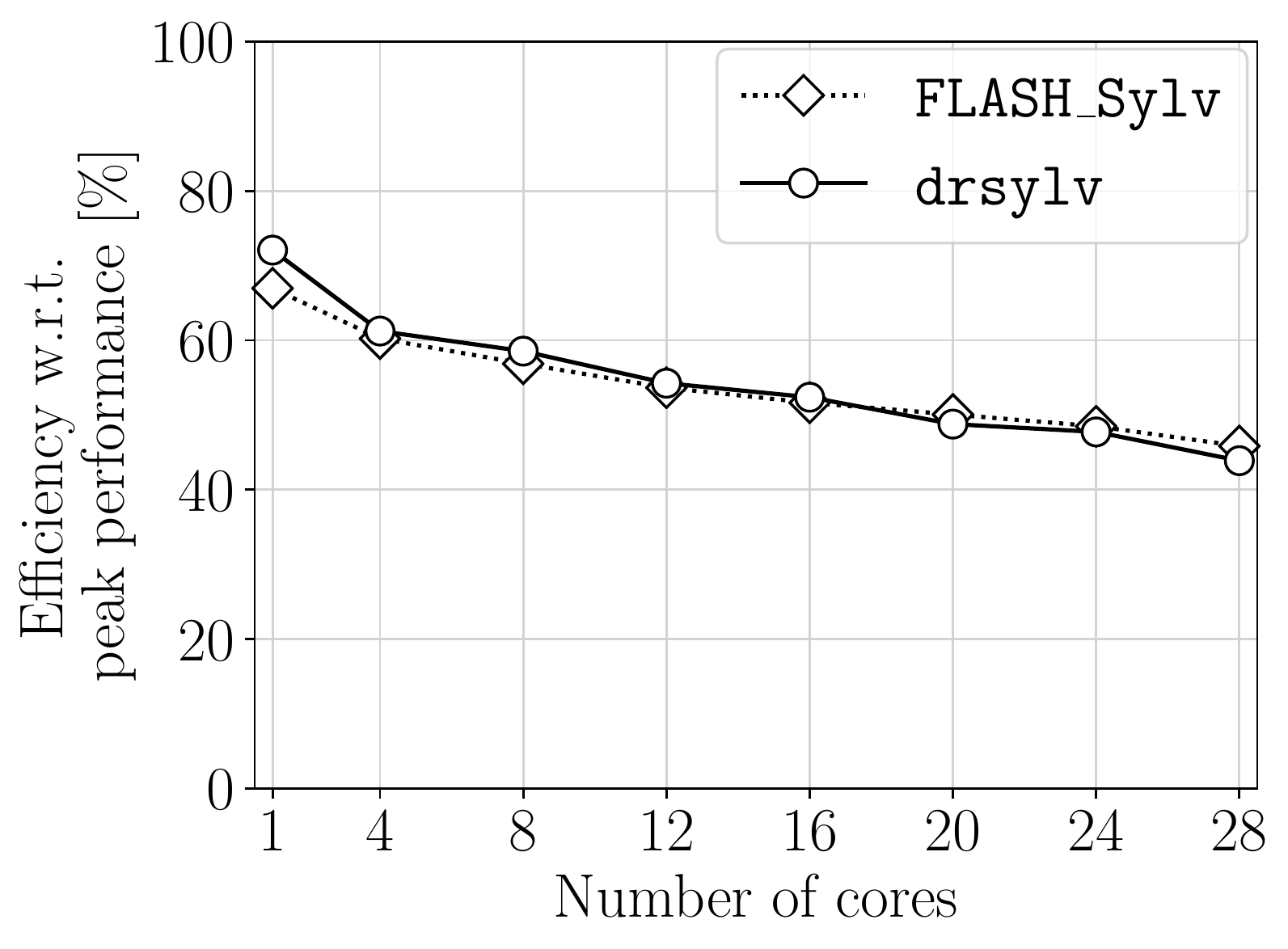}
    $m = n = 20000$
  \end{minipage}
  \caption{Strong scalability on systems that do not require scaling.}\label{fig:strongscaling}
\end{figure}

\paragraph{Strong Scalability without Numerical Scaling}
We examine the strong scalability of \texttt{FLASH\_Syl} and \texttt{drsylv} on one shared memory node. Perfect strong scalability manifests itself in a constant efficiency when the number of cores is increased for a fixed problem. Figure~\ref{fig:strongscaling} shows the results for $\mu = m$ and $\nu = n$. Robustness as implemented in \texttt{drsylv} does not hamper the scalability on systems that do not require scaling.

\paragraph{Cost of Robustness}
\begin{figure}[h!]
  \begin{minipage}{.5\textwidth}
  \includegraphics[width=\linewidth]{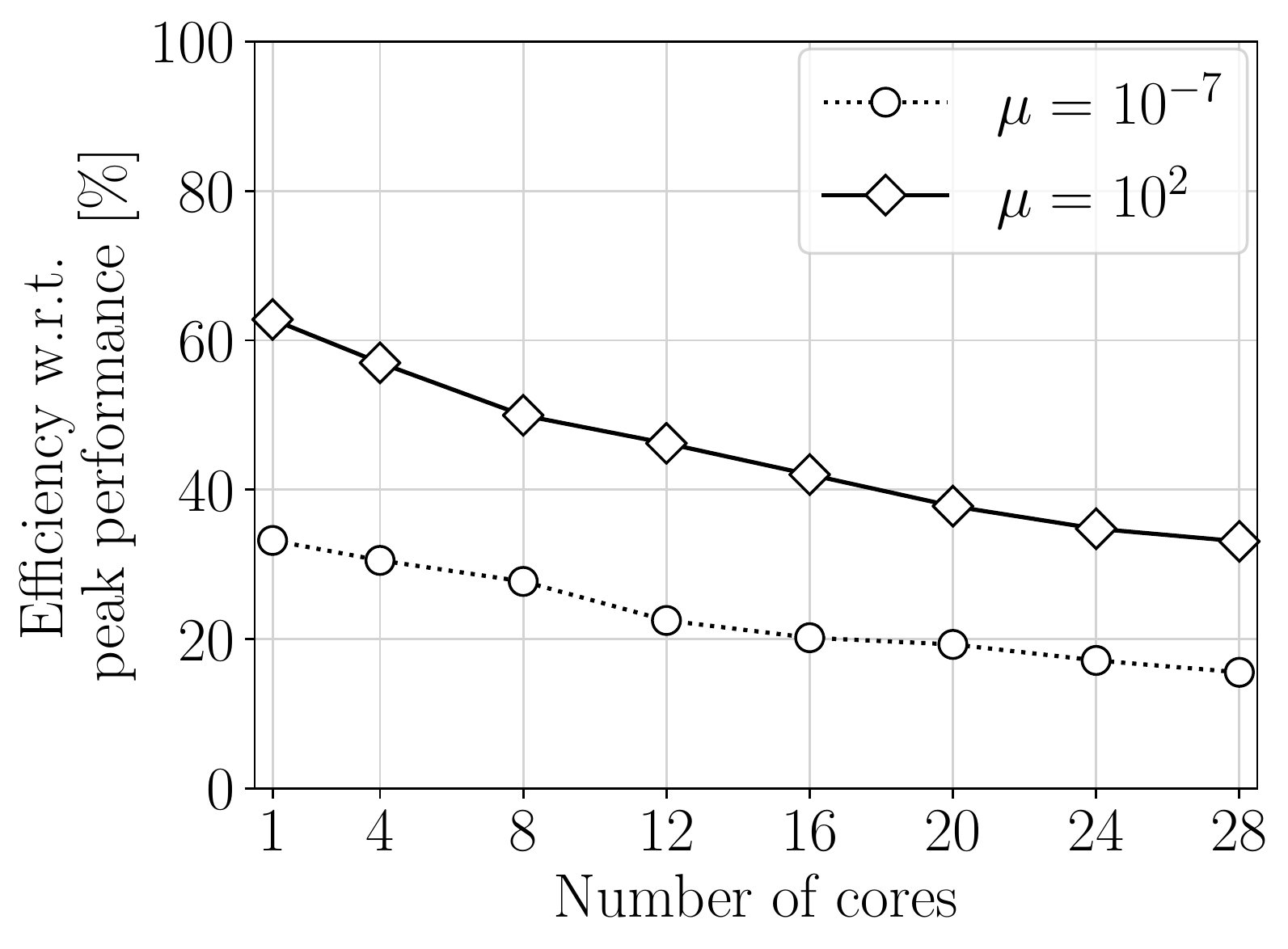}
  \end{minipage}
  \begin{minipage}{.5\textwidth}
  \includegraphics[width=\linewidth]{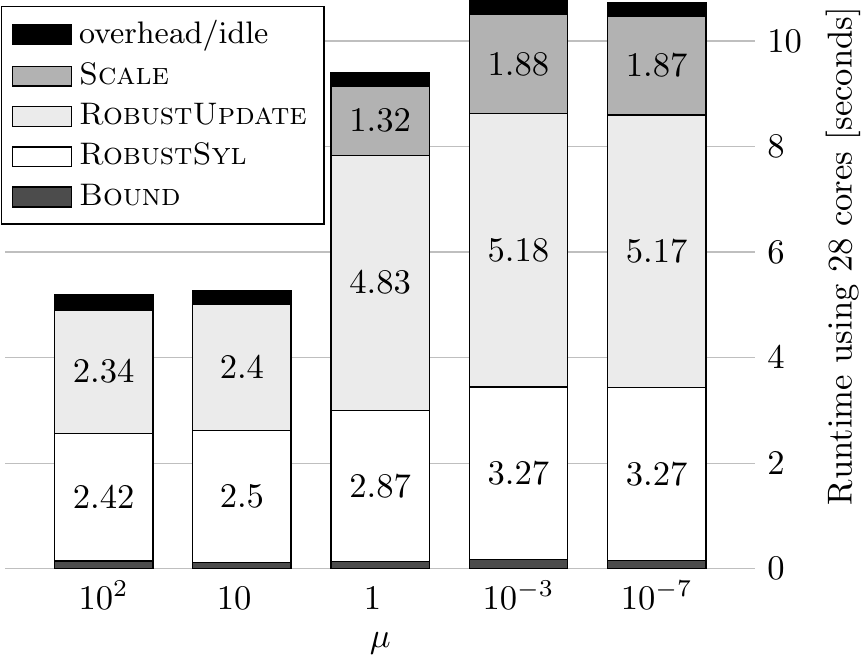}
  \end{minipage}
  \caption{Cost of robustness for $m = n = 10000$.}\label{fig:CostRobustness}
\end{figure}

Figure~\ref{fig:CostRobustness} (right) analyzes the cost of robustness. The amount of scaling necessary is controlled by fixing $\nu = 10^{-2}$ and varying $\mu$. The experiments with $\mu = 10^2, 10$ do not require scaling. The choice $\mu=1$ triggers scaling for a few tiles. Frequent scaling is required for $\mu = 10^{-3}, 10^{-7}$. The cost of \textsc{RobustUpdate} increases when some of the three input tiles require scaling. In the worst case all three tiles have to be rescaled in every update. Figure~\ref{fig:CostRobustness} (left) shows that despite of robustness a decent fraction of the peak performance is reached.

\section{Conclusion}\label{sec:conclusion}

The solution of the triangular Sylvester equation is a central step in the Bartels-Stewart algorithm. During the backward substitution, the components of the solution can exhibit growth, possibly exceeding the representable floating-point range. This paper introduced a task-parallel solver with overflow protection. By adding overflow protection, task-parallel solvers can now solve the same set of problems that is solvable with LAPACK.

The numerical experiments revealed that the overhead of overflow protection is negligible when scaling is not needed. Hence, the non-robust solver offers no real advantage over our robust solver. When scaling is necessary, our robust algorithm automatically applies scaling to prevent overflow. While scaling increases the runtime, it guarantees a representable result. The computed solution can be evaluated in the context of the user's application. This certainty cannot be achieved with a non-robust algorithm.

\paragraph*{Acknowledgements}\label{sec:acknowledgements}
The authors thank the research group for their support. This project has received funding from the European Union's Horizon 2020 research and innovation programme under grant agreement No 671633. Support was received by eSSENCE, a collaborative e-Science programme funded by the Swedish Government via the Swedish Research Council
(VR).


\bibliographystyle{splncs04}
\bibliography{sylvester}

\begin{thebibliography}{1}
\providecommand{\url}[1]{\texttt{#1}}
\providecommand{\urlprefix}{URL }
\providecommand{\doi}[1]{https://doi.org/#1}

\bibitem{bartels1972}
Bartels, R.H., Stewart, G.W.: {Solution of the Matrix Equation $AX+XB=C$}.
  Communications of the ACM  \textbf{15}(9),  820--826 (1972)

\bibitem{golub1996matrix}
Golub, G.H., Van~Loan, C.F.: {Matrix Computations}. John Hopkins University
  Press, 3rd edn. (1996)

\bibitem{higham2002accuracy}
Higham, N.J.: {Accuracy and Stability of Numerical Algorithms}. SIAM, 2nd edn.
  (2002)

\bibitem{jonsson2002part1}
Jonsson, I., K{\aa}gstr{\"o}m, B.: {Recursive blocked algorithms for solving
  triangular systems -- Part I: One-sided and coupled Sylvester-type matrix
  equations}. ACM TOMS  \textbf{28}(4),  392--415 (2002)

\bibitem{mikkelsen2017nlafet}
{Kjelgaard~Mikkelsen}, C.C., Karlsson, L.: {Robust Solution of Triangular
  Linear Systems} (2017), {NLAFET Working Note 9}

\bibitem{mikkelsen2017blocked}
Mikkelsen, C.C.K., Karlsson, L.: {Blocked Algorithms for Robust Solution of
  Triangular Linear Systems}. In: International Conference on Parallel
  Processing and Applied Mathematics. pp. 68--78. Springer (2017)

\bibitem{quintana2003formal}
Quintana-Ort{\'\i}, E.S., Van De~Geijn, R.A.: {Formal Derivation of Algorithms:
  The Triangular Sylvester Equation}. ACM TOMS  \textbf{29}(2),  218--243
  (2003)

\bibitem{simoncini2016computational}
Simoncini, V.: {Computational Methods for Linear Matrix Equations}. SIAM Review
   \textbf{58}(3),  377--441 (2016)

\bibitem{libflame}
Van~Zee, F.G.: {libflame: The Complete Reference (version 5.1.0-56)}

\end{thebibliography}

\end{document}